\documentclass{svproc}
\usepackage{graphicx}
\usepackage{amssymb}
\usepackage{amsmath}
\usepackage{dcolumn}
\usepackage{bm}
\usepackage{epsfig}
\usepackage{setspace}
\usepackage[caption=false]{subfig}
\usepackage{ulem}

\usepackage{url}

\begin{document}
\mainmatter              
\title{Time-evolution of net-baryon density fluctuations across the QCD critical region}
\titlerunning{Time-evolution of critical net-baryon density fluctuations}  
%
\author{Marcus Bluhm\inst{1,2} \and Marlene Nahrgang\inst{1,2}}
\authorrunning{M.~Bluhm and M.~Nahrgang} 
%
\tocauthor{Marcus Bluhm, Marlene Nahrgang}
\institute{$^1$~SUBATECH UMR 6457 (IMT Atlantique, Universit\'e de Nantes, \\IN2P3/CNRS), 4 rue Alfred Kastler, 44307 Nantes, France\\
$^2$~ExtreMe Matter Institute EMMI, GSI, Planckstr. 1, 64291 Darmstadt, Germany\\
}

\maketitle              

\begin{abstract}
We investigate the role of a finite surface tension during the time-evolution of fluctuations in the net-baryon density. The systems in this study undergo a temperature evolution across the phase transition in the critical region of the QCD phase diagram. The occuring non-equilibrium effects are discussed. 

\end{abstract}
\section{Introduction}
The search for the conjectured critical point associated with the deconfinement and chiral phase transition in the QCD phase diagram has received tremendously increasing attention in recent years, both theoretically and experimentally. In order to identify potential signatures of the critical point in experimental data a firm understanding of the dynamics of critical fluctuations during the evolution of matter created in heavy-ion collisions is crucial. For the net-baryon density $n_B$ as the slow critical mode~\cite{Hohenberg:1977ym} this dynamics is governed by diffusion processes. The time-evolution of fluctuations in $n_B$, which are expected to be enhanced in the critical region~\cite{Stephanov:2008qz,Asakawa:2009aj}, can therefore be modeled by a stochastic diffusion equation. Numerical simulations of the diffusion of critical fluctuations both on the crossover~\cite{Sakaida:2017rtj,Nahrgang:2017hkh,Nahrgang:2018afz} and first-order phase transition sides of the critical point have been performed recently for one spatial dimension. In these studies, non-equilibrium effects~\cite{Berdnikov:1999ph,Nahrgang:2011mg,Kitazawa:2013bta,Mukherjee:2015swa} such as retardation or critical slowing down have been observed quantitatively. 

In this talk, we study the time evolution of net-baryon density fluctuations  across the phase transition in the crossover domain near the critical point. We look at the local variance, i.e. the variance of event-by-event fluctuations over the size of a fluid cell. In continuum it is related to the zero-distance value of the two-point correlation function.
%
%
We numerically solve the stochastic diffusion equation in the form 
\begin{multline}
 \partial_t n_B(x,t) = \frac{D}{n_c} \left(m^2 \nabla_x^2 n_B - K\nabla_x^4 n_B\right) + \sqrt{2Dn_c/A}\,\nabla_x\zeta_x(x,t) \\ + D\nabla_x^2\left(\frac{\lambda_3}{n_c^2}\, (\Delta n_B)^2 + \frac{\lambda_4}{n_c^3}\, (\Delta n_B)^3 + \frac{\lambda_6}{n_c^5}\, (\Delta n_B)^5\right) \,.
\label{eq:stochdiff}
\end{multline}
Here, $\Delta n_B=n_B-n_c$ with a critical density of $n_c=1/(3$~fm$^3)$. 
Equation~\eqref{eq:stochdiff} describes the dynamics of the critical fluctuations in one spatial, the longitudinal, dimension which is not coupled to the physics in the transverse area $A$. Accordingly, the covariance of the white noise reads $\langle\zeta_x(x,t),\zeta_x(x',t')\rangle=\delta(x-x')\delta(t-t')$. This ensures that the long-time equilibrium distribution of $n_B$ is governed by the free energy of the system in agreement with the fluctuation-dissipation theorem.

The particular form of Eq.~\eqref{eq:stochdiff} is based on a free energy functional which contains apart from a Gaussian mass term, where $m^2$ is inversely connected with the correlation length $\xi$, also a kinetic energy term with surface tension $K$ and non-linear coupling terms which are proportional to $\lambda_i$. This is valid in the vicinity of the critical point and may still be improved by including further regular contributions. The dependence of the coupling coefficients on $\xi$ is defined by universality~\cite{Hohenberg:1977ym} through the mapping of the $3$-dimensional Ising spin model onto an effective potential~\cite{Tsypin:1997zz} and the dependence of $\xi$ on temperature $T$ and baryo-chemical potential $\mu_B$ in QCD can be determined by a matching to the susceptibility of the spin model scaling equation of state~\cite{Guida:1996ep} as explained in~\cite{Bluhm:2016byc,Bluhm:2016trm}. The numerical results presented in this work depend on the values of the dimensionless couplings $\tilde\lambda_i$ and the diffusion coefficient $D$. 

The stochastic diffusion equation~\eqref{eq:stochdiff} allows us to contrast three distinct physical cases: the Gauss model with $K=\lambda_i=0$ and $m^2\ne 0$ which was studied in detail in~\cite{Nahrgang:2017hkh}, the Gauss$+$surface model for which also $K\ne 0$, and the Ginzburg-Landau model as considered in~\cite{Nahrgang:2018afz} for which we exemplarily choose $\tilde\lambda_3=1$, $\tilde\lambda_4=10$ and $\tilde\lambda_6=3$ for the dimensionless couplings. In the numerical implementation we apply a semi-implicit scheme for the Ginzburg-Landau model and an implicit scheme for the other two. Exact net-baryon number conservation is ensured by imposing periodic boundary conditions. 
\section{Time evolution of the variance}
To study the time evolution of fluctuation observables such as the local variance $\sigma^2$, we consider a system of temporally evolving but spatially constant $T$ which is otherwise static, i.e. of a fixed finite size $L=20$~fm. The temperature cools down following $T(\tau)=T_0 \left(\tau_0/\tau\right)$ where we start from $T_0=0.5$~GeV with equilibrated initial conditions at $\tau_0=1$~fm$/$c. Correspondingly, $D(\tau)=D(\tau_0)T(\tau)/T_0$. At $\tau-\tau_0=2.33$~fm$/$c the pseudo-critical temperature is reached. 

\begin{figure}[t]
 \centering
 \includegraphics{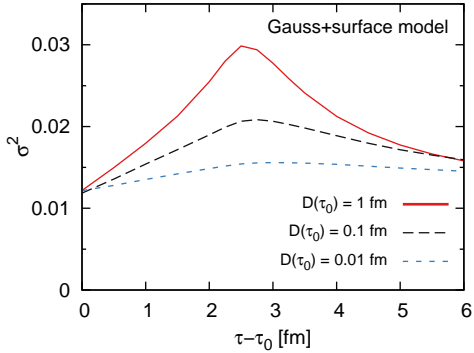}
 \vspace{-1mm}
 \caption{Comparison of the time-evolution of the local variance $\sigma^2$ for the Gauss$+$surface model for different diffusion coefficients $D/\textrm{fm}=1,\,0.1,\,0.01$ at $\tau_0=1$~fm$/$c. The critical temperature $T_c$ is reached at  $\tau-\tau_0=2.33$~fm$/$c.
 \label{fig:1}}
\end{figure}

Let us first focus on the Gauss$+$surface model. For this model analytic expressions for the long-time equilibrium value of the local variance at fixed $T$ can be derived both in the continuum and in discretized space-time. In the latter case the result depends on the employed resolution $\Delta x=L/N_x$ with $N_x$ considered lattice sites and reads $\sigma^2=n_c^2\,e^{-\Delta x \sqrt{m^2/K}}/(2A\sqrt{m^2K})$. In equilibrium, physical observables such as $\sigma^2$ are independent of the diffusion coefficient $D$. As the temperature cools down, the system falls out of equilibrium as a consequence of this dynamics and the time-evolution of $\sigma^2$ becomes sensitive to the value of $D$. This is illustrated in Fig.~\ref{fig:1}. One observes that with decreasing diffusion coefficient finite-range fluctuations are not formed fast enough and as a consequence their magnitude is reduced during the time evolution. Similar observations have been reported for the Gauss model in~\cite{Nahrgang:2017hkh}. Here, even for $D(\tau_0)=1$~fm the maximum of $\sigma^2$ is reduced by a factor of $2$ compared to the equilibrium value at $T_c$ due to non-equilibrium effects. The temporal evolution of $T$ is much larger than the diffusion rate. Moreover, this maximum is shifted towards later times, i.e.~smaller $T$ than $T_c$, where this retardation effect becomes stronger with decreasing $D$.

\begin{figure}[t]
 \centering
 \includegraphics{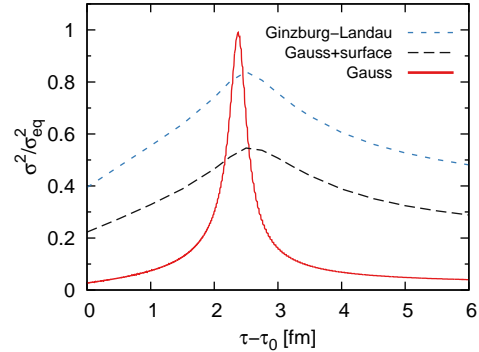}
 \vspace{-1mm}
 \caption{Comparison of the time-evolution of the local variance $\sigma^2$ scaled by its equilibrium value $\sigma^2_{\textrm{eq}}$ at $T=T_c$ for different forms of the stochastic diffusion equation (Gauss, Gauss$+$surface and Ginzburg-Landau models) and $D(\tau_0)=1$~fm. The equilibrium value has been calculated for a static box of length $L=20$~fm at constant temperature in the long-time limit.
 \label{fig:2}}
\end{figure}

In Fig.~\ref{fig:2} we contrast the impact of the time-evolution of $T$ on the behavior of the local variance for the three different physical models. One observes that the non-equilibrium effects are larger once a finite surface tension is included because it takes more time to build up the finite correlation length associated with $K$. The local variance in the Gauss model reaches almost its equilibrium value at $T=T_c$, see~\cite{Nahrgang:2017hkh}, while it is significantly reduced once $K\neq0$. Moreover, it is interesting to point out another effect of finite $K$: the local variance at a temperature away from $T_c$ is not suppressed as much as it is the case in the Gauss model. The retardation effect is larger for the models with $K\neq 0$, as the maximal value of the local variance is shifted to $T<T_c$. 

\section{Conclusions and outlook}
In this work, we have outlined the importance of finite-range correlations associated with the presence of a surface tension $K$ in models describing the diffusion of net-baryon density fluctuations. The time needed to build up these correlations can be important and of the same order as the cooling time, which leads to important non-equilibrium effects. 

In future work, the expansion of the system, e.g. in a Bjorken geometry, will be added and second-order relaxation equations will be applied in order to guarantee causality and prepare a future embedding in full fluid dynamical simulations of heavy-ion collisions. 
\paragraph{Acknowledgments}
The authors acknowledge the support of the program "Etoiles montantes en Pays de la Loire 2017". The work was in parts supported by the ExtreMe Matter Institute (EMMI) at the GSI Helmholtzzentrum f\"ur Schwerionenforschung, Darmstadt, Germany. The authors thank T.~Sch\"afer, S.~Bass, M.~Kitazawa and the members of the BEST Topical Collaboration for many stimulating discussions. 
%
%

%
\end{document}